\documentclass[lettersize,journal]{IEEEtran}
\usepackage{amsmath,amsfonts}
\usepackage{algorithmic}
\usepackage{algorithm}
\usepackage{array}
\usepackage{amsthm}
\usepackage{textcomp}
\usepackage{stfloats}
\usepackage{url}
\usepackage{verbatim}
\usepackage{graphicx}
\usepackage{subcaption}
\usepackage{epstopdf}
\usepackage{cite}
\usepackage{cleveref}
\Crefname{figure}{Fig.}{Figures}
\usepackage{amssymb,amsthm}
 
\newtheorem{remark}{Remark} 

\newtheorem{theorem}{Theorem}

\begin{document}

\title{Performance Analysis of BEM-based Channel Estimation for OTFS with Hardware Impairments}
\author{Haowei Wu, \textit{Member, IEEE}, Huanyu Chen, Qihao Peng, Qu Luo, and Jinglan Ou
 \thanks{Haowei Wu, Huanyu Chen, and Jinglan Ou are with the School of Microelectronics and Communication Engineering, Chongqing University, Chongqing 400044, China, and also with the Chongqing Key Laboratory of Space Information Network and Intelligent Information Fusion. (e-mail: wuhaowei@cqu.edu.cn; chenhuanyu77@163.com; oujinglan@cqu.edu.cn). Qihao Peng and Qu luo are with 5G and 6G Innovation Centre, Institute for Communication Systems (ICS) of University of Surrey, Guildford, GU2 7XH, UK. (e-mail: \{q.peng, q.u.luo\}@surrey.ac.uk). (\it{Corresponding Author: Qihao Peng}.)}.}

\markboth{Journal of \LaTeX\ Class Files,~Vol.~14, No.~8, August~2021}%
{Shell \MakeLowercase{\textit{et al.}}: A Sample Article Using IEEEtran.cls for IEEE Journals}


\maketitle

\begin{abstract}
This letter studies the low-complexity channel estimation for orthogonal time frequency space (OTFS) in the presence of hardware impairments. Firstly, to tackle the computational complexity of channel estimation, the basis expansion model (BEM) is utilized. Then, the mean square error (MSE) of the estimated channel is theoretically derived, revealing the effects of hardware impairments on channel estimation. Based on the estimated channel, the minimum mean square error (MMSE) detector is adopted to analyze the impacts of imperfect hardware on the bit error rate (BER). Finally, the numerical results validate the correctness of our theoretical analysis of the MSE for channel estimation and lower bound of the BER, and also demonstrate that even minor hardware impairments can significantly degrade the performance of the OTFS system.
\end{abstract}

\begin{IEEEkeywords}
Orthogonal time frequency space (OTFS), hardware impairments, minimum mean square error (MMSE), basis expansion model (BEM), bit error rate (BER).
\end{IEEEkeywords}

\section{Introduction}
\IEEEPARstart{T}he legacy orthogonal frequency division multiplexing (OFDM) techniques have been widely used in 4G and 5G networks, thanks to its advantages (e.g., low-complexity implementation, capability of inter-symbol interference (ISI) mitigation) in linear time-invariant channels\cite{ref1,ref2}. However, its sensitivity to Doppler shift leads to significant performance degradation in high-speed mobile scenarios. To tackle this issue, the orthogonal time frequency space (OTFS) is proposed for efficient transmission over time-variant channels. Through a two-dimensional (2D) transformation, OTFS converts the time-varying channel in the time-frequency (TF) domain to quasi-static channel in the delay-Doppler (DD) domain and exhibits sparsity \cite{ref2,ref3}. Due to its inherent 2D modulation characteristics, OTFS faces challenges related to accurate channel estimation.

To date, extensive efforts have been devoted to investigating the channel estimation of OTFS \cite{ref4,ref5,ref6}. In \cite{ref4}, an embedded pilot scheme and a threshold-based channel estimation method were proposed, which require exhaustive search for the optimal parameters. With this in mind, the authors of \cite{ref5,ref6} decomposed time-varying channels using the basis expansion model (BEM) for continuous Doppler spread channels under the ideal hardware platform, which significantly reduces the channel estimation complexity. However, \cite{ref4,ref5,ref6} did not account for practical hardware impairments, which are crucial in real-world scenarios.  In \cite{ref7}, the author analyzed the performance of OTFS in the presence of hardware impairments such as in phase and quadrature (IQ) imbalance. Later on, the authors of \cite{ref8} investigated hardware effects, such as IQ imbalance and direct current (DC) offset, on the system performance and proposed methods for estimating and compensating the associated impairments. More recently, the authors in \cite{ref9} proposed an autoencoder-based OTFS scheme for eliminating the effects of hardware impairments and improving error performance. It is worth noting that existing studies have generally investigated channel estimation and hardware impairments independently, and the impact of hardware impairments on channel estimation and BER remains unknown.

In this letter, we propose a BEM-based low-complexity channel estimation framework for OTFS, meticulously designed to address the challenges in the presence of  hardware implementations and fractional Doppler. Firstly, the impact of hardware impairments on channel estimation is analyzed through the rigorously theoretical derivation of the MSE. Subsequently, the average BER performance of OTFS considering hardware impairments is analyzed using the MMSE detector. Finally, the simulation results demonstrated that hardware impairments significantly degraded the error performance and validated our theoretical analysis.

The remainder of this letter is organized as follows. Section \ref{sec_2} describes the OTFS system model based on the BEM considering hardware impairments. Based on the received pilot, the channel estimation,  equalization, and BER analysis are given in Section \ref{sec_3}. The numerical results are presented in Section \ref{sec_4}. The conclusions are drawn in Section \ref{sec_5}.

\section{System Model}\label{sec_2} 
\begin{figure}[!h]
\centering
\includegraphics[width=0.5\textwidth]{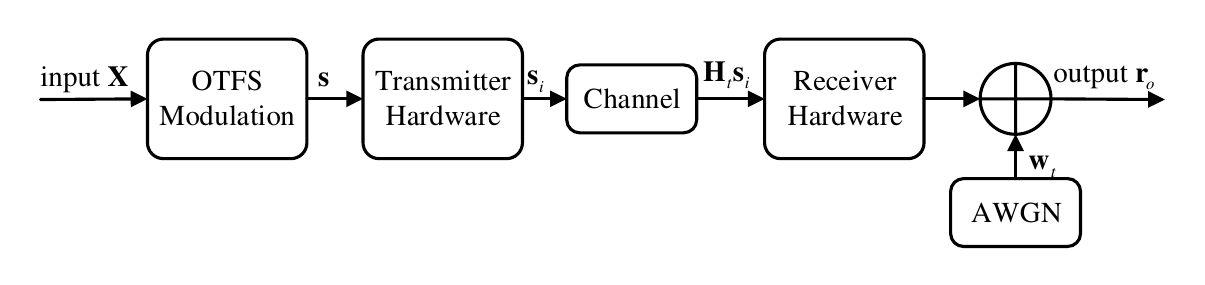}
\caption{Hardware Impairment OTFS System Model.}
\label{fig_1}
\end{figure}

\subsection{Hardware Impairment OTFS Signal Model}
The block diagram of the OTFS system in the presence of hardware impairments is shown in \Cref{fig_1}. The frame of OTFS information in the DD domain is indicated by ${\mathbf{X}} = {{\mathbf{X}}_d} + {{\mathbf{X}}_p} \in {\mathbb{C}^{M \times N}}$, where $M$ and $N$ denote the sizes of the delay and Doppler dimension grid, respectively. ${{\mathbf{X}}_d}$ and ${{\mathbf{X}}_p}$ denote the data and the pilot, respectively. Then, ${{\mathbf{X}}}$ is converted into the one-dimensional time-domain signal by ${\mathbf{s}} = {\text{vec}}\left( {{{\mathbf{G}}_{tx}}{\mathbf{XF}}_N^H} \right) \in {\mathbb{C}^{NM \times 1}}$, where $\text{vec}\left(  \cdot  \right)$ represents vectorization operations, ${\mathbf{F}}_N^H \in \mathbb{C}^{N \times N}$ denotes the normalized $N$ points inverse discrete Fourier transform (IDFT) matrix, and the superscripts $H$ denotes complex conjugate transpose. ${{\mathbf{G}}_{tx}} \in {\mathbb{C}^{M \times M}}$ represents transmitter pulse shaping. In this letter, the rectangular window function is used, i.e, ${{\mathbf{G}}_{tx}} = {{\mathbf{I}}_M}$. To mitigate the effects of multiple paths, the cyclic prefix (CP) of length ${L_{CP}} \geqslant L$ is appended to ${\mathbf{s}}$, where $L$ is the maximum delay spread. Considering the non-ideal hardware, the signals at the transmitter and receiver are respectively given by \cite{ref10,ref11}
\begin{equation}\label{eq1}
    {{\mathbf{s}}_i} = \sqrt {{\xi _i}} {\mathbf{s}} + {{\mathbf{z}}_i},
\end{equation}
\begin{equation}
    \begin{aligned}\label{eq2}
  {{\mathbf{r}}_o} &= \sqrt {{\xi _o}{\xi _i}} {{\mathbf{H}}_t}{\mathbf{s}} + \sqrt {{\xi _o}} {{\mathbf{H}}_t}{{\mathbf{z}}_i} + {{\mathbf{w}}_t} + {{\mathbf{z}}_o}, \\ 
    \end{aligned}
\end{equation}
where ${{\xi _i}}$ and ${{\xi _o}}$ represent the transmitter and receiver hardware quality factors, respectively. ${{\mathbf{z}}_i} \sim \mathcal{C}\mathcal{N}\left( {{\bf 0},\left( {1 - {\xi _i}} \right){ \mathbb E}\left\{ {{{\left| {{\mathbf s}} \right|}^2}} \right\} \mathbf{I}_{NM}} \right)$, ${{\mathbf{z}}_o} \sim \mathcal{C}\mathcal{N}\left( {{\bf 0},\left( {1 - {\xi _o}} \right)\mathbb{E}\left( {{{\left| {{\mathbf{H}}_t}{{\mathbf{s}}_i} \right|}^2}} \right){{{\mathbf I}}_{NM}}} \right)$ represent the additive distortion noise terms caused by the hardware at the transmitter and receiver sides, respectively. Both ${{\mathbf{z}}_i}$ and ${{\mathbf{z}}_o}$ follow the complex Gaussian distribution with $\bf{0}$ mean, but with variances given by $\left( {1 - {\xi _i}} \right)\mathbb{E}\left\{ {{{\left| {\mathbf{s}} \right|}^2}} \right\}{{\mathbf{I}}_{NM}}$ and $\left( {1 - {\xi _o}} \right)\mathbb{E}\left( {{{\left| {{\mathbf{H}}_t}{{\mathbf{s}}_i} \right|}^2}} \right){{\mathbf{I}}_{NM}}$, respectively. $\mathbb{E}\left\{  \cdot  \right\}$ denotes expectation operation, ${{\mathbf{H}}_t} \in {\mathbb{C}^{NM \times NM}}$ denotes the channel matrix in the time domain, and ${{\mathbf{w}}_t} \sim \mathcal{C}\mathcal{N}\left( {{\bf 0},\sigma _w^2{{\mathbf{I}}_{NM}}} \right)$ is additive Gaussian white noise (AWGN) with mean $\bf{0}$ and variance $\sigma _w^2{{\mathbf{I}}_{NM}}$. 

The received signal in the DD domain is
\begin{equation}\begin{aligned}\label{eq3}
  {\mathbf{y}} &= {\mathbf{F}}{{\mathbf{r}}_0} \hfill = \sqrt {{\xi _o}{\xi _i}} {{\mathbf{H}}_{DD}}{\mathbf{x}} + {{\mathbf{z}}_n} \hfill, \\ 
\end{aligned}\end{equation}
where ${\mathbf{F}} = \left( {{{\mathbf{F}}_N} \otimes {{\mathbf{I}}_M}} \right)\in {\mathbb{C}^{NM \times NM}}$ represents the transformation matrix, and ${{\mathbf{H}}_{DD}} = {\mathbf{F}}{{\mathbf{H}}_t}{{\mathbf{F}}^H}$ is the channel matrix in the DD domain. Define ${{\mathbf{z}}_n} = \sqrt {{\xi _o}} {\mathbf{F}}{{\mathbf{H}}_t}{{\mathbf{z}}_i} + {\mathbf{F}}{{\mathbf{w}}_t} + {\mathbf{F}}{{\mathbf{z}}_o}$, then $\mathbf z_n$ can be treated as non-ideal hardware and random channel noise. 
\subsection{GCE-BEM Channel Model}
The time-varying channel is modeled using the generalized complex exponential BEM (GCE-BEM), and the channel coefficients of the ${l'}$th path at time $t$ can be denoted as
\begin{equation}\label{eq4}
    h\left[ {t,l'} \right] = \sum\limits_{q = 0}^{Q_L} {{c_q}\left[ {l'} \right]{e^{j{w_q}t}}}  + {e_{\bmod }}\left( {t,l'} \right),\forall l' \in \{ 0, \cdots ,L\} ,
\end{equation}
where ${Q_L} \ge 2\left\lceil {\frac{{RN{f_{\max }}}}{{\Delta f}}} \right\rceil $  denotes the BEM order, ${{c_q}\left[ {l'} \right]}$ the ${q}$th BEM basis function coefficient corresponding to the ${l'}$th path, ${w_q} = \frac{{2\pi }}{{MNR}}\left( {q - \left\lceil {\frac{Q_L}{2}} \right\rceil } \right)$ is the ${q}$th BEM modeling frequency, $\left\lceil  \cdot  \right\rceil $ represents the ceiling operator, ${R}$ denotes frequency resolution factor, and ${e_{\bmod }}\left( {t,l'} \right)$ denotes the modeling error\cite{ref5}. In the case of $R = 1$, (\ref{eq4}) takes its simplest form but also exhibits the largest modeling error\cite{ref5}. We refer (\ref{eq4}) with $R = 1$ as the simplest complex exponential BEM (CE-BEM), and denote ${Q_L} \ge{Q_S} = 2\left\lceil {\frac{{N{f_{\max }}}}{{\Delta f}}} \right\rceil $. Considering the trade-off between the complexity and modeling error, $R=2$ is employed, and the corresponding channel in time domain can be expressed as
\begin{equation}\label{eq5}
{{\mathbf{H}}_t} = \sum\limits_{q = 0}^{Q_L} {{\text{diag}}\left( {{{\mathbf{b}}_q}} \right){\mathbf{F}}_{MN}^H{\text{diag}}\left( {{{\mathbf{F}}_{MN \times L}}{{\mathbf{c}}_q}} \right){{\mathbf{F}}_{MN}}}  + {{\mathbf{E}}_{\bmod }},\end{equation}
where ${{\mathbf{b}}_q} = {\left[ {{e^{j{w_q}0}},{e^{j{w_q}1}},...,{e^{j{w_q}\left( {NM - 1} \right)}}} \right]^T}$ refers to the $q$th BEM basis function vector, the $(\cdot)^T$ denotes transpose, ${{\mathbf{c}}_q} = {\left[ {{c_q}\left[ 0 \right],{c_q}\left[ 1 \right],...,{c_q}\left[ L \right]} \right]^T}$ denotes the $q$th coefficient vector, ${{\mathbf{E}}_{\bmod }}$ denotes the channel modeling error matrix, and ${{{\bf{F}}_{MN \times L}}}$ is the first columns $L$ of the $MN$ points discrete Fourier transform (DFT) matrix. Considering a rich scattering environment \cite{ref6}, the classical Jakes’ channel model with a U-shaped Doppler spectrum is employed. Based on this model, we define the covariance matrix of the $l'$th path ${{\mathbf{h}}_{l'}} = \left[ {{{\mathbf{h}}_{l'}}\left[ 0 \right],{{\mathbf{h}}_{l'}}\left[ 1 \right],...,{{\mathbf{h}}_{l'}}\left[ {NM - 1} \right]} \right]$ as ${{\mathbf{R}}_{{{\mathbf{h}}_{l'}}}}$, and its \([m,n]\)th element can be represented as
\begin{equation}\begin{aligned}\label{eq6}
    &{{\mathbf{R}}_{{{\mathbf{h}}_{l'}}}}\left[ {m,n} \right] = \mathbb{E}\left\{ {{{\mathbf{h}}_{l'}}\left[ m \right]{\mathbf{h}}_{l'}^ * \left[ n \right]} \right\} \\ 
   = &\sigma _{l'}^2{J_0}\left( {2\pi {f_{\max }}\left| {m - n} \right|{T_s}} \right), m,n \in \left[ {0,NM - 1} \right], 
\end{aligned}\end{equation}
where $T_s$ is the sampling interval, $\sigma _{l'}^2$ denotes the channel power of the $l'$th path with $\sum\limits_{l' = 0}^L {\sigma _{l'}^2 = 1} $, and ${J_0}\left(  \cdot  \right)$ represents the first type of zeroth-order Bessel function. ${f_{\max }} = \frac{{{f_c}{v_{\max }}}}{c}$ is the maximum Doppler shift, where ${f_c}$, ${v_{\max }}$ and $c$ is the carrier frequency, the speed of user, and the speed of light, respectively. 

\section{BEM-OTFS Receiver}\label{sec_3}
\begin{figure}[!h]
\centering
\includegraphics[width=0.5\textwidth]{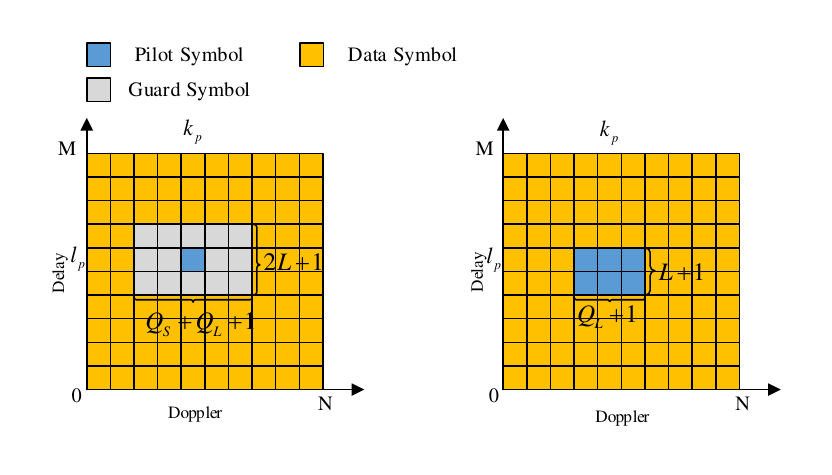}
\caption{OTFS System Pilot Patten.}
\label{fig_2}
\end{figure}

\subsection{Pilot Design}
As illustrated in \Cref{fig_2}, the nonzero single pilot is located at $\left[ {{l_p},{k_p}} \right]$th square, and the number of guard symbols along the delay and Doppler dimensions are $2L+1$ and ${Q_s} + {Q_L} +1$, respectively. As observed, this approach sacrifices part of the pilot overhead but facilitates direct estimation of the GCE-BEM coefficients, thereby reducing the computational complexity \cite{ref5}. The pilot overhead is defined as $\lambda  = \frac{{\left( {2L + 1} \right)\left( {{Q_s} + {Q_L} + 1} \right)}}{{NM}}$. Therefore, the transmitted symbol matrix in the DD domain can be expressed as
\begin{equation}\label{eq7}
{\mathbf{X}}\left[ {l,k} \right] = \left\{ {\begin{array}{*{20}{c}}\!\!\!\!\!\!\!\!\!\!\!
  {{x_p},}&{k = {k_p},l = {l_p}} \\ 
  \!\!\!\!\!\!\!\!\!{0,}&\begin{gathered}
  \!\!\!\!\!\!\!\!\!\!l \in \left[ {{l_p} - L,{l_p} + L} \right], \hfill \\
 \!\!\!\!\!\!\!\!\!\!k \in \left[ {{k_p} - \frac{{{Q_S} + {Q_L}}}{2},{k_p} + \frac{{{Q_S} + {Q_L}}}{2}} \right] \hfill \\ 
\end{gathered}  \\ 
  \!\!{{x_d}\left[ {l,k} \right],}&{{\text{otherwise}}} 
\end{array}} \right.,\end{equation}
where ${{x_p}}$ and ${{x_d}\left[ {l,k} \right]}$ denote the pilot and data, respectively. Owing to the multiple paths and Doppler, the pilot signal diffuses into the $k \in \left[ {{k_p} - \frac{{{Q_L}}}{2},{k_p} + \frac{{{Q_L}}}{2}} \right]$, $l \in \left[ {{l_p},{l_p} + L} \right]$,  as the blue region depicted in \Cref{fig_2}. Based on this received pilot signal, the channel can be estimated.
\subsection{Channel Estimation}
In this subsection, we present the proposed channel estimation scheme in the presence of hardware impairments. Substituting (\ref{eq5}) into (\ref{eq3}), we have
\begin{equation}\label{eq8}
{\mathbf{y}} = \sqrt {{\xi _o}{\xi _i}} {{\mathbf{A}}_p}{\mathbf{c}} + \sqrt {{\xi _o}{\xi _i}} {{\mathbf{A}}_d}{\mathbf{c}} + \sqrt {{\xi _o}{\xi _i}} {{\mathbf{z}}_{\text{mod}}} + {{\mathbf{z}}_n},
\end{equation}
where ${{\mathbf{A}}_p} = \left[ {{{\mathbf{A}}_{p,0}},{{\mathbf{A}}_{p,1}},...,{{\mathbf{A}}_{p,{Q_L}}}} \right] \in {\mathbb{C}^{NM \times \left( {{Q_L} + 1} \right)\left( {L + 1} \right)}}$ is a deterministic matrix related to the pilot. ${{\mathbf{A}}_{p,q}} = {{\mathbf{V}}_q}{\text{diag}}\left( {{{\mathbf{F}}_{MN}}{{\mathbf{s}}_p}} \right){{\mathbf{F}}_{MN \times L}} \in {\mathbb{C}^{NM \times \left( {L + 1} \right)}}$ represent pilot matrix, where ${{\mathbf{V}}_q} = {\mathbf{F}}{\text{diag}}\left( {{{\mathbf{b}}_q}} \right){\mathbf{F}}_{MN}^H \in {\mathbb{C}^{NM \times NM}}$ is related to the $q$th BEM basis function vector, ${{\mathbf{s}}_p}$ represents the pilot signal vector transmitted in the time domain. ${{\mathbf{A}}_d} = \left[ {{{\mathbf{A}}_{d,0}},{{\mathbf{A}}_{d,1}},...,{{\mathbf{A}}_{d,{Q_L}}}} \right]\in {\mathbb{C}^{NM \times \left( {{Q_L} + 1} \right)\left( {L + 1} \right)}}$ is data-dependent and is a random matrix. ${{\mathbf{A}}_{d,q}} = {{\mathbf{V}}_q}{\text{diag}}\left( {{{\mathbf{F}}_{MN}}{{\mathbf{s}}_d}} \right){{\mathbf{F}}_{MN \times L}}\in {\mathbb{C}^{NM \times \left( {L + 1} \right)}}$ represent data matrix, where ${{\mathbf{s}}_d}$ is the data vector transmitted in the time domain. ${\mathbf{c}} = {\left[ {{\mathbf{c}}_0^T,{\mathbf{c}}_1^T,...,{\mathbf{c}}_{{Q_L}}^T} \right]^T} = {\left( {{\mathbf{B}} \otimes {{\mathbf{I}}_{L + 1}}} \right)^\dag }{\mathbf{h}}\in {\mathbb{C}^{\left({Q_L+1} \right)\times \left( {L + 1} \right)}}$ denotes BEM coefficients vector, where ${\mathbf{B}} = \left[ {{{\mathbf{b}}_0},{{\mathbf{b}}_1},...,{{\mathbf{b}}_{{Q_L}}}} \right]$ consists of BEM basis functions vectors, $\dag$ represents pseudo inverse of a vector/matrix. ${\mathbf{h}} = \left[ {{{\mathbf{h}}_0}\left[ 0 \right],{{\mathbf{h}}_1}\left[ 0 \right],...,{{\mathbf{h}}_L}\left[ {NM - 1} \right]} \right]$ denotes time-varying channel impulse response (CIR) vector of length $MN(L+1)$, which is assumed to follow the complex Gaussian random vector with zero mean. ${{\mathbf{z}}_{\text{mod} }} = {\mathbf{F}}{{\mathbf{E}}_{\text{mod} }}{\mathbf{s}}$ is the vector form of the modeling error. 

The portion of the received signal associated with the pilot is used to estimate the channel and is given by ${\mathbf{Ey}} = \sqrt {{\xi _o}{\xi _i}} {\mathbf{E}}{{\mathbf{A}}_p}{\mathbf{c}} + {\mathbf{z}}$, where ${\mathbf{E}}$ denotes the matrix that can extract the pilot signal from the received signal with ${\mathbf{E}}{{\mathbf{E}}^H} = {{\mathbf{I}}_{\left( {{Q_L} + 1} \right)\left( {L + 1} \right)}}$.  By treating the data interference as a part of noise, we have ${\mathbf{z}} = \sqrt {{\xi _o}{\xi _i}} {\mathbf{E}}{{\mathbf{A}}_d}{\mathbf{c}} + \sqrt {{\xi _o}{\xi _i}} {\mathbf{E}}{{\mathbf{z}}_{\text{mod} }} + {\mathbf{E}}{{\bf{z}}_n}$. Then, we have the following theorem.

\begin{theorem}\label{theorem1}
Based on the provided signal, using the MMSE estimator, the BEM coefficient is estimated as
\begin{equation}\label{eq9}
{\mathbf{\hat c}}\! =\!\!\sqrt {{\xi _o}{\xi _i}} {{\mathbf{R}}_{\mathbf{c}}}{\mathbf{A}}_p^H{{\mathbf{E}}^H}{\left( {{\xi _o}{\xi _i}{\mathbf{E}}\left( {{{\mathbf{A}}_p}{{\mathbf{R}}_{\mathbf{c}}}{\mathbf{A}}_p^H + {{\mathbf{R}}_{\mathbf{z}}}} \right){{\mathbf{E}}^H}} \right)^{ - 1}}\left( {{\mathbf{Ey}}} \right),
\end{equation}
where ${{\mathbf{R}}_{\mathbf{c}}} = {\left( {{\mathbf{B}} \otimes {{\mathbf{I}}_{L + 1}}} \right)^\dag }{{\mathbf{R}}_{{\mathbf{hh}}}}{\left( {{{\mathbf{B}}^H} \otimes {{\mathbf{I}}_{L + 1}}} \right)^\dag }$ and ${{\mathbf{R}}_{{\mathbf{hh}}}}$ denotes the covariance matrix of the BEM coefficient vectors and the CIR vectors, respectively. $\otimes$ denotes matrix Kronecker product. ${{\mathbf{R}}_{\mathbf{z}}} = \mathbb{E}\left\{ {{\mathbf{z}}{{\mathbf{z}}^H}} \right\}$ represents the correlation matrix of noise plus interference vector, which is given at the top of the next page. In (\ref{eq13}), ${\bf{\Phi }} = {{\bf{\Phi }}_d} + {{\bf{\Phi }}_p}$, and ${{\bf{\Phi }}_d}= {\mathbb E}\left\{ {{{\bf{x}}_d}{\bf{x}}_d^H} \right\}$, ${{\bf{\Phi }}_p} = {\mathbb E}\left\{ {{{\bf{x}}_p}{\bf{x}}_p^H} \right\}$ denote the correlation matrix of data and pilot symbols, respectively.
 ${{\mathbf{R}}_{{q_1},{q_2}}}= \left( {{{\mathbf{F}}_{MN}}{{\mathbf{F}}^H}{{\mathbf{\Phi }}_d}{\mathbf{FF}}_{_{MN}}^H} \right) \odot \left( {{{\mathbf{F}}_{MN \times L}}{{\mathbf{R}}_{{{\mathbf{c}}_{{q_1}}},{{\mathbf{c}}_{q2}}}}{\mathbf{F}}_{_{MN \times L}}^H} \right)$, where ${{\bf{R}}_{{{\bf{c}}_{{q_1}}},{{\bf{c}}_{{q_2}}}}} = {\mathbb E}\left\{ {{{\bf{c}}_{{q_1}}}{\bf{c}}_{{q_2}}^H} \right\}$, and $ \odot $ is Hadamard product.
 
Based on the estimated \(\mathbf{\hat c}\), the estimated channel matrix can be given by
\begin{equation}\label{eq10}
{{\mathbf{\hat H}}_t} = \sum\limits_{q = 0}^{{Q_L}} {{\text{diag}}\left( {{{\mathbf{b}}_q}} \right){\mathbf{F}}_{MN}^H{\text{diag}}\left( {{{\mathbf{F}}_{MN \times L}}{{{\mathbf{\hat c}}}_q}} \right){{\mathbf{F}}_{MN}}}.
\end{equation}
The covariance matrix of BEM estimation error can be expressed as
\begin{equation}\label{eq11}
\begin{aligned}
  {{\mathbf{R}}_{{\mathbf{\tilde c}}}}= {\mathbf{K}}{{\mathbf{R}}_{\mathbf{c}}}{{\mathbf{K}}^H} + {\xi _o}{\xi _i}{\mathbf{DE}}{{\mathbf{R}}_{\mathbf{z}}}{{\mathbf{E}}^H}{{\mathbf{D}}^H} , 
\end{aligned}
\end{equation}
where \({\mathbf{K}}\) and \({\mathbf{D}}\) are given by
\begin{equation}\label{eq_12}
    \begin{split}
        {\mathbf{K}} &= \left( {{{\mathbf{I}}_{\left( {{Q_L} + 1} \right)\left( {L + 1} \right)}} - {\xi _o}{\xi _i}{\mathbf{DE}}{{\mathbf{A}}_p}} \right)\\
        {\mathbf{D}} &= {{\mathbf{R}}_{\mathbf{c}}}{\mathbf{A}}_p^{\text{H}}{{\mathbf{E}}^{\text{H}}}{\left( {{\xi _o}{\xi _i}{\mathbf{E}}{{\mathbf{A}}_p}{{\mathbf{R}}_{\mathbf{c}}}{\mathbf{A}}_p^{\text{H}}{{\mathbf{E}}^{\text{H}}} + {\mathbf{E}}{{\mathbf{R}}_{\mathbf{z}}}{{\mathbf{E}}^{\text{H}}}} \right)^{ - 1}}.
    \end{split}
\end{equation}

Then, the corresponding MSE can be expressed as
\begin{equation}\label{eq12}
   {\text{MS}}{{\text{E}}_{{\text{CE}}}} = \frac{{{\text{Trace}}\left\{ {{\mathbf{\Omega }}{{\mathbf{R}}_{{\mathbf{\tilde c}}}}{{\mathbf{\Omega }}^H}} \right\} + \sum\limits_{l' = 0}^L {{\text{Trace}}\left\{ {{\mathbf{G}}{{\mathbf{R}}_{{\mathbf{h}},l'}}} \right\}} }}{{MN(L + 1)}},
\end{equation}
where ${\mathbf{\Omega }} = \left( {{{\mathbf{I}}_{L + 1}} \otimes {\mathbf{B}}} \right){\mathbf{P}}$, ${\mathbf{P}}$ is the permutation matrix, with ${{\mathbf{P}}^H}{\mathbf{P}} = {{\bf{I}}_{\left( {{Q_L} + 1} \right)\left( {L + 1} \right)}}$, ${\bf{G}} = {{\bf{I}}_{NM}} - {\bf{B}}{\left( {{{\bf{B}}^H}{\bf{B}}} \right)^{ - 1}}{{\bf{B}}^H}$. ${\text{Trace}}\left\{  \cdot  \right\}$ denotes the trace of a matrix.

\textit{Proof: See Appendix \ref{appendix_1}.} \(\hfill\blacksquare\)
\end{theorem}
 \begin{figure*}[ht]\begin{equation}\begin{aligned}\label{eq13}
{{\mathbf{R}}_{\mathbf{z}}} = {\xi _o}{\xi _i}{\mathbf{E}}\left( {\sum\limits_{{q_1} = 0}^{{Q_L}} {\sum\limits_{{q_2} = 0}^{{Q_L}} {{{\mathbf{V}}_{{q_1}}}{{\mathbf{R}}_{{q_1},{q_2}}}{\mathbf{V}}_{{q_2}}^H} }  + {\mathbf{F}}\sum\limits_{l' = 0}^L {\left( {{{\mathbf{\Pi }}^{l'}}{\mathbf{F\Phi }}{{\left( {{{\mathbf{\Pi }}^{l'}}{\mathbf{F}}} \right)}^H}} \right) \odot {{\mathbf{R}}_{{{\mathbf{e}}_{\rm{mod,}l'}}}}} {{\mathbf{F}}^H}} \right){{\mathbf{E}}^H} + \left( {{\xi _o}\sigma _{{{\mathbf{z}}_i}}^2 + \sigma _{{{\mathbf{w}}}}^2 + \sigma _{{{\mathbf{z}}_o}}^2} \right){\mathbf{I}}.
\end{aligned}\end{equation}\end{figure*}
\begin{remark}
For an ideal OTFS communication system without hardware impairments, the parameters in (\ref{eq13}) simplified as ${{\bf{R}}_{\bf{z}}} = {\bf{EF}}\sum\limits_{l' = 0}^L {\left( {{{\bf{\Pi }}^{l'}}{\bf{F\Phi }}{{\left( {{{\bf{\Pi }}^{l'}}{\bf{F}}} \right)}^H}} \right) \odot {{\bf{R}}_{{{\bf{e}}_{\text{mod,}l'}}}}} {{\bf{F}}^H}{{\bf{E}}^H} + {\bf{E}}\sum\limits_{{q_1} = 0}^{{Q_L}} {\sum\limits_{{q_2} = 0}^{{Q_L}} {{{\bf{V}}_{{q_1}}}{{\bf{R}}_{{q_1},{q_2}}}{\bf{V}}_{{q_2}}^H} } {{\bf{E}}^H} + \left( {\sigma _{{{\bf{w}}}}^2} \right){{\bf{I}}_{\left( {{Q_L} + 1} \right)\left( {L + 1} \right)}}$, This indicates that the system is free from amplitude attenuation and distortion noise introduced by non-ideal hardware. Consequently, the findings can be considered a lower bound on the final estimated performance.
\end{remark}
\begin{remark}
As the quality factor of the hardware, i.e., ${\xi _o} $, ${\xi _i}$ and $X$, are progressively decreased, the attenuation of the useful signal is increased while introducing greater additive distortion noise, thereby making (\ref{eq12}) larger. In the worst case, the quality factor at either end of the transceiver is reduced to zero, the numerator of (\ref{eq12}) is rewritten as ${\text{Trace}}\left\{ {{\bf{\Omega }}{{\bf{R}}_{\bf{c}}}{{\bf{\Omega }}^H}} \right\}+ \sum\limits_{l' = 0}^L {{\text{Trace}}\left\{ {{\mathbf{G}}{{\mathbf{R}}_{{\mathbf{h}},l'}}} \right\}} $, at which point the useful signal component is completely filtered and the system is rendered useless.
\end{remark}
\subsection{BER Analysis}
Using (\ref{eq10}), the pilot interference is extracted from the received signal, written as
\begin{equation}\label{eq14}
{{\mathbf{\hat r}}_o} = {{\mathbf{r}}_o} - \sqrt {{\xi _o}{\xi _i}} {{\mathbf{\hat H}}_t}{{\mathbf{s}}_p} = \sqrt {{\xi _o}{\xi _i}} {{\mathbf{\hat H}}_t}{{\mathbf{s}}_d} + {\mathbf{n}},
\end{equation}
where ${\mathbf{n}} = \sqrt {{\xi _o}{\xi _i}} {{\mathbf{\tilde H}}_t}{\mathbf{s}} + \sqrt {{\xi _o}{\xi _i}} {{\mathbf{z}}_{\text{mod} }} + \sqrt {{\xi _o}} {{\mathbf{H}}_t}{{\mathbf{z}}_i} + {{\mathbf{w}}_t} + {{\mathbf{z}}_o}$ denotes the estimation error plus noise vector, and ${{\mathbf{\tilde H}}_t} = {{\mathbf{H}}_t} - {{\mathbf{\hat H}}_t}$ is the channel estimation error. 

Then, the MMSE detector was implemented in time domain
\begin{equation}\label{eq15}
    {{\mathbf{\hat s}}_d} = {{\mathbf{G}}_t}{{\mathbf{\hat r}}_o},
\end{equation}
where ${{\mathbf{G}}_t} = \sqrt {{\xi _o}{\xi _i}} {{\mathbf{R}}_{{{\mathbf{s}}_d}}}{\mathbf{\hat H}}_t^H{\left( {{\xi _o}{\xi _i}{{\mathbf{\hat H}}_t}{{\mathbf{R}}_{{{\mathbf{s}}_d}}}{\mathbf{\hat H}}_t^H + {{\mathbf{R}}_{\mathbf{n}}}} \right)^{ - 1}}$ denotes the detecting matrix in the time domain, ${{\mathbf{\hat s}}_d}$ is the estimated time domain data, ${{\mathbf{R}}_{{{\mathbf{s}}_d}}} = \mathbb{E}\left\{ {{{\mathbf{s}}_d}{\mathbf{s}}_d^H} \right\}$and ${{\mathbf{R}}_{\mathbf{n}}} = \mathbb{E}\left\{ {{\mathbf{n}}{{\mathbf{n}}^H}} \right\}$. Then, we have the following theorem.

\begin{theorem}\label{theorem2}
Based on the MMSE detector, the theoretical average BER of a quadrature amplitude modulation (QAM) mapping using Gray coding can be lower bounded by
\begin{equation}\label{eq16}
\begin{aligned}
    {P_{OTFS}} &= \frac{1}{{{N_{{\text{num}}}}}}\sum\limits_{i = 0}^{{N_{{\text{num}}}}} {{a_M}{\text{erfc}}\left( {\sqrt {{b_M}{\text{SINR}}\left[ i \right]} } \right)}\\
    &\geqslant {a_M}{\text{erfc}}\left( {\sqrt {\frac{{{b_M}\sum\limits_{i = 1}^{{N_{{\text{num}}}}} {\frac{{{\mathbf{T}}\left[ {i,i} \right]}}{N}} }}{{1 - \sum\limits_{i = 1}^{{N_{{\text{num}}}}} {\frac{{{\mathbf{T}}\left[ {i,i} \right]}}{N}} }}} } \right),
\end{aligned}
\end{equation}
where ${\text{erfc}}\left( x \right) = \frac{2}{{\sqrt \pi  }}\int_x^\infty  {{e^{ - {t^2}}}dt} $ is the Complementary Error Function, and \(\mathbf{T}[i,i]\) is the \([i,i]\)th element of ${\mathbf{T}} = \sqrt {{\xi _o}{\xi _i}} {\mathbf{F}}{{\mathbf{G}}_t}{{\mathbf{H}}_t}{{\mathbf{F}}^H} \in {\mathbb{C}^{NM \times NM}}$, The parameters, ${{N_{{\text{num}}}}}$ denote the total number of data symbols in the DD domain, ${{a_M}}$ and ${{b_M}}$ are related to the type of modulation mapping \cite{ref13}. For 4-QAM, \({{a_M}}\) and \({{b_M}}\) are $\frac{1}{2}$.

\textit{Proof: see Appendix \ref{appendix_2}.} \(\hfill \blacksquare\)
\end{theorem}

\section{Simulation Results}\label{sec_4}
To characterize the impacts of hardware impairments on OTFS and validate the analytic derivation, unless stated, the following parameter settings are employed for all simulation. In this letter, we consider $M = 64,N = 16$, $L = 5$, ${f_c} = 4$ GHz, $\Delta f = 30$ kHz. The maximum moving speed is ${v_{\max }} = 500$ km/h, corresponding to a maximum Doppler shift ${f_{\max }} = \frac{{{f_c}{v_{\max }}}}{c} = 1852$ Hz. Therefore, the size of the guard interval in the pilot pattern are ${Q_s} = 2$ and ${Q_L} = 4$, respectively. The Jakes' Doppler spectrum is given by ${v_i} = {v_{\max }}\cos \left( \theta  \right)$, where $\theta $ follows a uniform distribution in the interval $\left[ { - \pi ,\pi } \right]$. The pilot overhead is given by $\lambda  = 7.5\% $, and we assume the data symbol and pilot has the same transmit power. Finally, 4-QAM is used.
\begin{figure}[!h]
\centering
\begin{subfigure}[b]{0.45\textwidth}
    \includegraphics[width=1\textwidth]{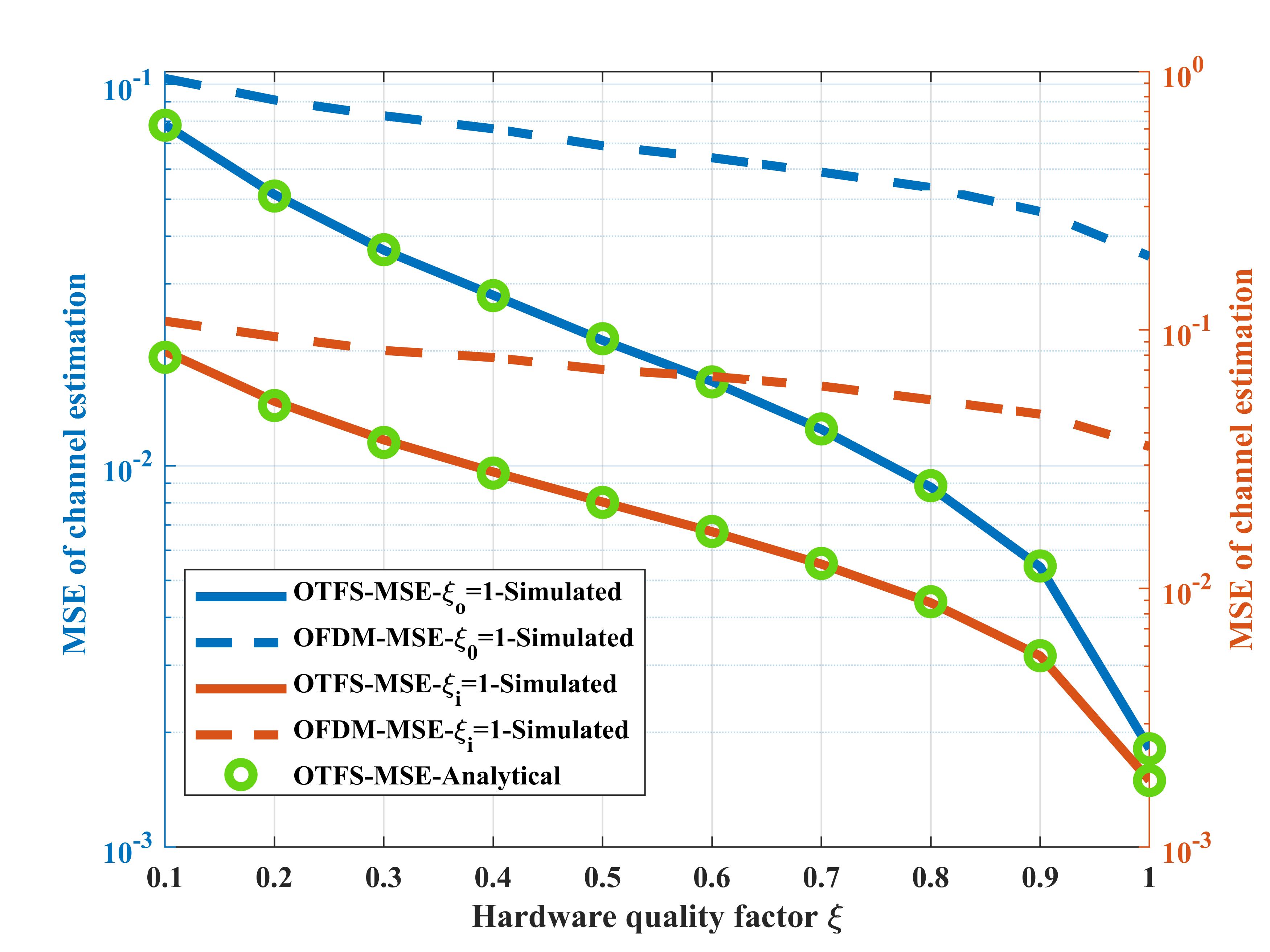}
    \caption{}
   \label{fig_3(a)}
\end{subfigure}
\begin{subfigure}[b]{0.45\textwidth}
    \includegraphics[width=1\textwidth]{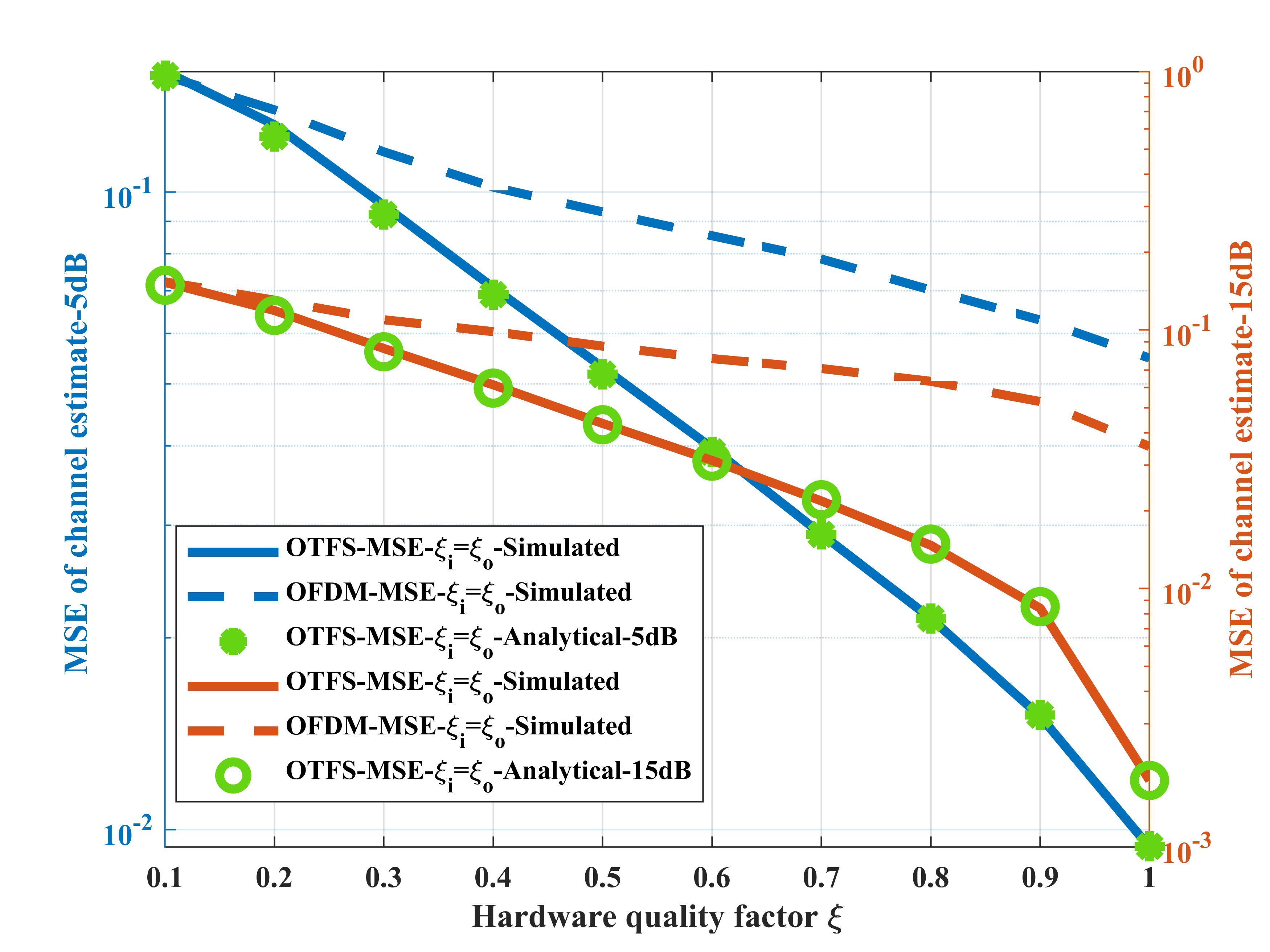}
     \caption{}
\label{fig_3(b)}
\end{subfigure}
\caption{MSE performance with different quality factors. (a) different ${\xi _i}$ or ${\xi _o} $ , with ${\text{SNR = 15dB}}$. (b) ${\xi _i}={\xi _o} $ for ${\text{SNR = 5dB}}$ and ${\text{SNR = 15dB}}$.}
\label{fig_3}
\end{figure}

\Cref{fig_3} shows the MSE performances of the proposed channel estimation with different quality factors. It is evident that the simulated MSE closely aligns with the theoretically derived MSE, substantiating the validity of the theoretical derivation. Secondly, the MSE performance of channel estimation for OTFS systems with hardware impairments exhibits a increase in volatility as the hardware quality factor rises. Furthermore, when the hardware quality factor is equivalent in the transceiver, the performance degradation is comparable. This is attributed to the fact that the distortion noise introduced is also comparable, as evidenced by (\ref{eq11}) and (\ref{eq13}), resulting in an approximate performance loss. Finally, the performance of the OTFS system is notably superior to that of OFDM in high-speed scenarios due to its resilience to Doppler shift.

\begin{figure}[!h]
\centering
\begin{subfigure}[b]{0.45\textwidth}
    \includegraphics[width=1\textwidth]{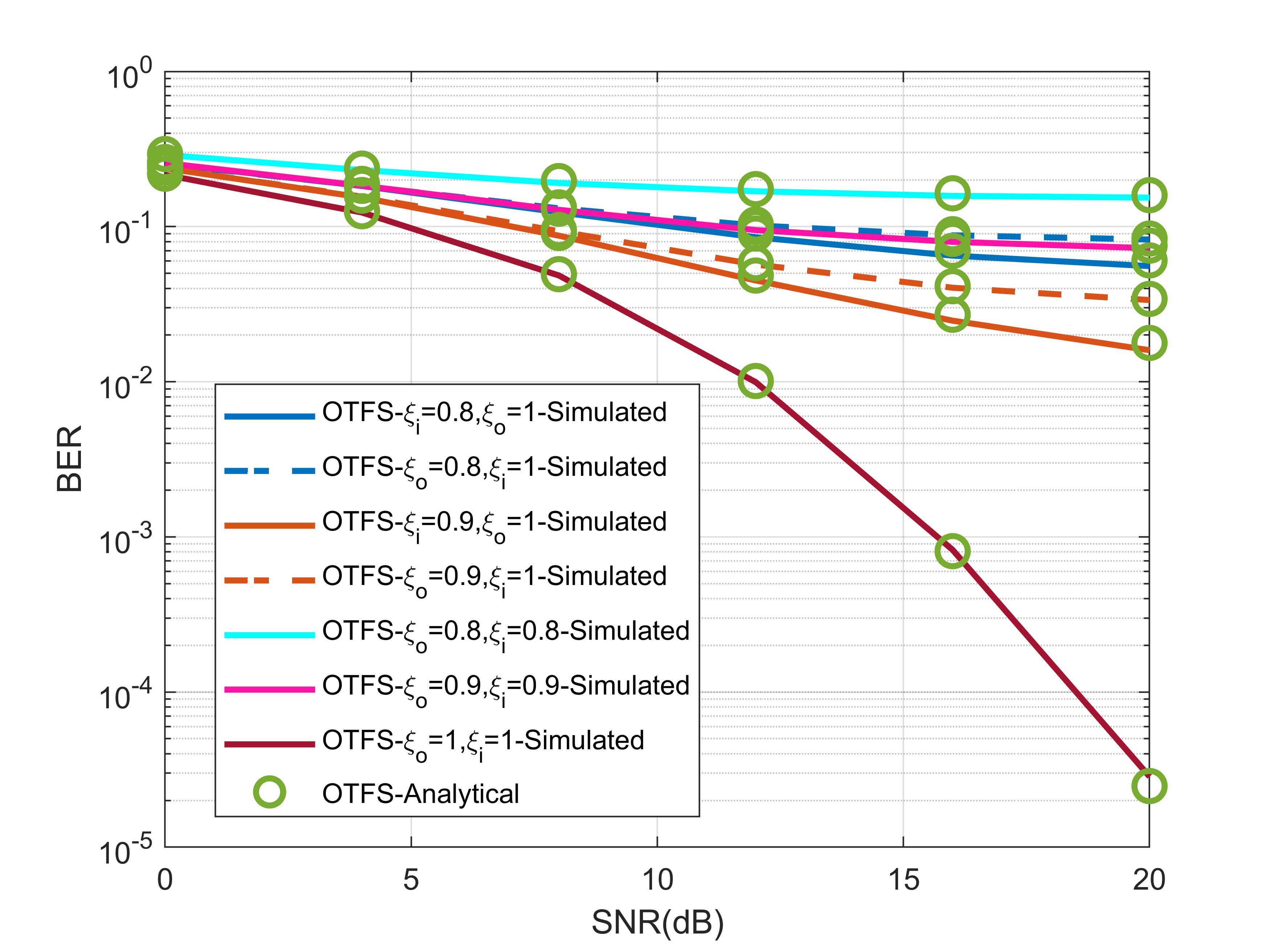}
    \caption{}
   \label{fig_4(a)}
\end{subfigure}
\begin{subfigure}[b]{0.45\textwidth}
    \includegraphics[width=1\textwidth]{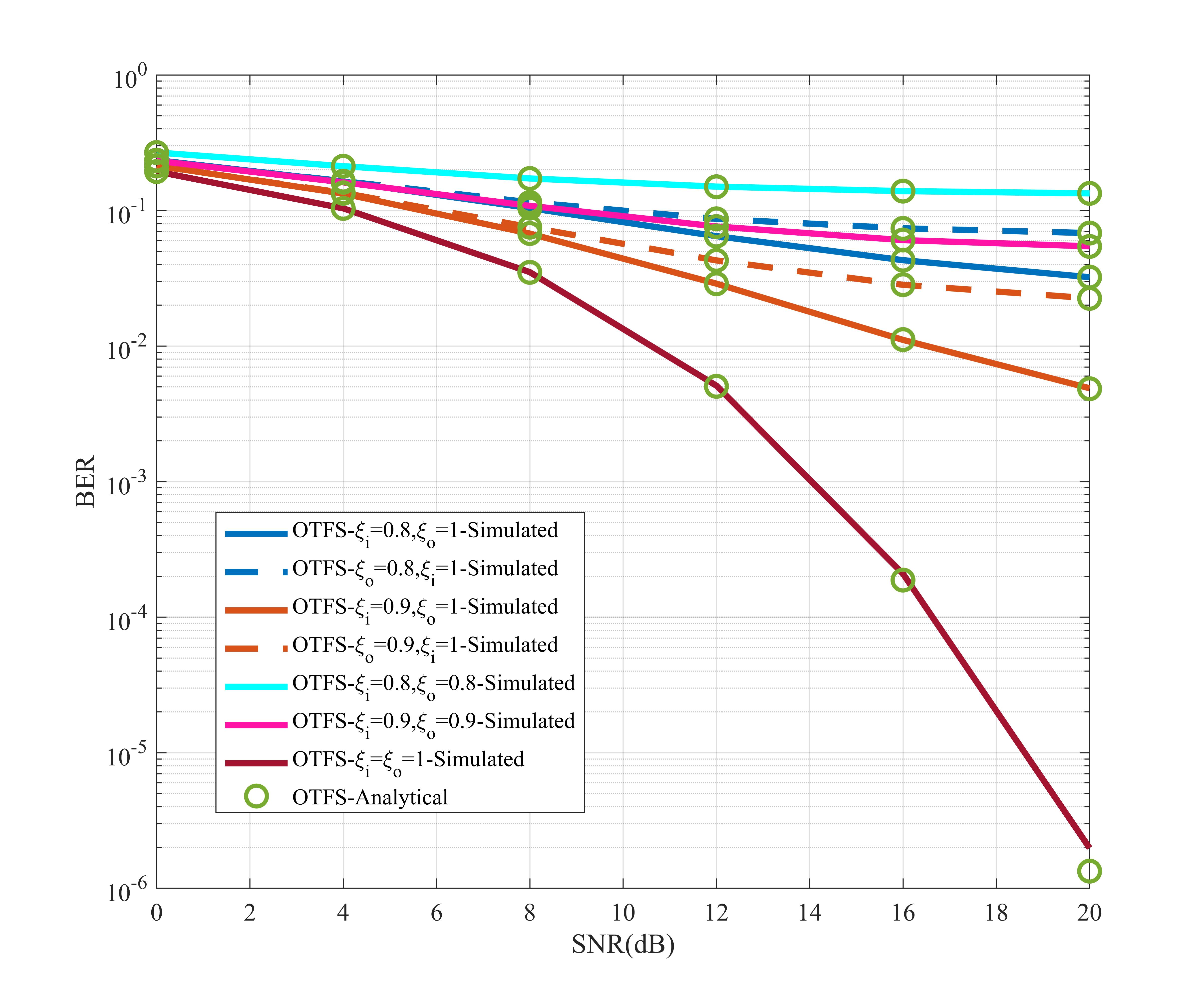}
     \caption{}
\label{fig_4(b)}
\end{subfigure}
\caption{BER performance with different quality factors. (a) Estimated channel. (b) Perfect channel.}
\label{fig_4}
\end{figure}
\Cref{fig_4} depicts the BER performance of the OTFS system in the presence of hardware impairments with different hardware parameters. It is evident that the simulated BER values closely match the theoretically analyzed BER lower bound throughout the SNR range, thus validating the precision of the analysis in Section \ref{sec_3}. As expected, the BER decreases as the increases of ${\xi_i}$ and ${\xi_o}$. The system achieves an ideal hardware configuration when the quality factor at the receiver and transmitter is increased to 1, i.e., ${\xi _i}={\xi _o}=1$. In low SNR regime, the BER performance under ideal hardware approaches that of non-ideal hardware, owing to the fact that AWGN is the predominant source of BER degradation. In contrast, in high SNR regime, a minor hardware impairment can have a substantial impact on BER performance, as hardware impairments are predominant, resulting in channel distortion and affecting the BER. This results in a significant reduction in the SINR of the OTFS signal and a corresponding drop in BER performance.
\section{Conclusion}\label{sec_5}
This letter characterized the performance of an OTFS system with hardware impairments for both receiver and transmitter based on BEM and analyzed the impact of hardware impairments on the MSE and BER performance of channel estimation. The simulation results validated the theoretical analysis and demonstrated that even minor hardware impairments can significantly degrade the error performance of OTFS systems.

{\appendix\subsection{Proof of Theorem \ref{theorem1}.}\label{appendix_1}
The interference plus noise vector for the estimation phase is ${\mathbf{z}} = \sqrt {{\xi _o}{\xi _i}} {\mathbf{E}}{{\mathbf{A}}_d}{\mathbf{c}} + \sqrt {{\xi _o}{\xi _i}} {\mathbf{E}}{{\mathbf{z}}_{\text{mod} }} + \sqrt {{\xi _o}} {\mathbf{EF}}{{\mathbf{H}}_t}{{\mathbf{z}}_i} + {\mathbf{EF}}{{\mathbf{w}}_t} + {\mathbf{EF}}{{\mathbf{z}}_o}$. Since ${{{\bf{z}}_i}}$, ${{{\bf{z}}_o}}$, and ${{{\bf{w}}_t}}$ are independent of each other with with zero mean, we have
\begin{equation}\label{eq17}\begin{aligned}
{\xi _o}\mathbb{E}\left\{ {{\mathbf{EF}}{{\mathbf{H}}_t}{{\mathbf{z}}_i}{\mathbf{z}}_i^H{\mathbf{H}}_t^H{{\mathbf{F}}^H}{{\mathbf{E}}^H}} \right\} &= {\xi _o}\sigma _i^2\sum\limits_{l' = 0}^L {\sigma _{l'}^2} {{\mathbf{I}}_{\left( {{Q_L} + 1} \right)\left( {L + 1} \right)}}\\
\mathbb{E}\left\{ {{\mathbf{EF}}{{\mathbf{w}}_t}{\mathbf{w}}_t^H{{\mathbf{F}}^H}{{\mathbf{E}}^H}} \right\} &= {\xi _o}\sigma _w^2{{\mathbf{I}}_{\left( {{Q_L} + 1} \right)\left( {L + 1} \right)}}\\
\mathbb{E}\left\{ {{\mathbf{EF}}{{\mathbf{z}}_o}{\mathbf{z}}_o^H{{\mathbf{F}}^H}{{\mathbf{E}}^H}} \right\} &= \sigma _{{{\mathbf{z}}_o}}^2{{\mathbf{I}}_{\left( {{Q_L} + 1} \right)\left( {L + 1} \right)}}.
\end{aligned}
\end{equation}
Similarly, as ${{\mathbf{c}}_q}$ and ${{{\mathbf{s}}_d}}$ are independent of each other, the covariance matrix of the data interference $\sqrt {{\xi _o}{\xi _i}}{{\mathbf{A}}_d}{\mathbf{c}} =\sqrt {{\xi _o}{\xi _i}} \sum\limits_{q = 0}^{{Q_L}} {{{\mathbf{V}}_q}{\text{diag}}\left\{ {{{\mathbf{F}}_{MN}}{{\mathbf{s}}_d}} \right\}{{\mathbf{F}}_{MN}}{{\mathbf{c}}_q}}$ can be derived as
\begin{equation}\label{eq18}
{\xi _o}{\xi _i}\mathbb{E}\left\{ {{{\mathbf{A}}_d}{\mathbf{c}}{{\mathbf{c}}^H}{\mathbf{A}}_d^H} \right\} = {\xi _o}{\xi _i}\sum\limits_{{q_1} = 0}^Q {\sum\limits_{{q_2} = 0}^Q {{{\mathbf{V}}_{{q_1}}}{{\mathbf{R}}_{{q_1},{q_2}}}{\mathbf{V}}_{{q_2}}^H} },
\end{equation}
where ${{\mathbf{R}}_{{q_1},{q_2}}}$ are given by
\begin{equation}\label{eq19}
    \begin{aligned}
   &\mathbb{E}\left\{ {{\text{diag}}\left\{ {{{\mathbf{F}}_{MN}}{{\mathbf{s}}_d}} \right\}{{\mathbf{F}}_{MN \times L}}{{\mathbf{c}}_{{q_1}}}{\mathbf{c}}_{{q_2}}^H{\mathbf{F}}_{_{MN \times L}}^H{\text{diag}}\left\{ {{\mathbf{s}}_d^H{\mathbf{F}}_{MN}^H} \right\}} \right\} \\ 
   &= \mathbb{E}\left\{ {{{\mathbf{F}}_{MN}}{{\mathbf{s}}_d}{\mathbf{s}}_d^H{\mathbf{F}}_{MN}^H} \right\} \odot \left( {{{\mathbf{F}}_{MN \times L}}\mathbb{E}\left\{ {{{\mathbf{c}}_{{q_1}}}{\mathbf{c}}_{{q_2}}^H} \right\}{\mathbf{F}}_{_{MN \times L}}^H} \right) \\ 
   &= \left( {{{\mathbf{F}}_{MN}}{{\mathbf{F}}^H}{{\mathbf{\Phi }}_d}{\mathbf{FF}}_{_{MN}}^H} \right) \odot \left( {{{\mathbf{F}}_{MN \times L}}{{\mathbf{R}}_{{{\mathbf{c}}_{{q_1}}},{{\mathbf{c}}_{q2}}}}{\mathbf{F}}_{_{MN \times L}}^H} \right).
    \end{aligned}
\end{equation}
${{\mathbf{\Phi }}_d} = \mathbb{E}\left\{ {{{\mathbf{x}}_d}{\mathbf{x}}_d^H} \right\}{\text{ = diag}}\left\{ {\mathbf{\bf{\gamma} }} \right\}$ represent the data power matrix in DD domain, and the elements of ${\mathbf{\gamma }} \in {\mathbb{R}^{NM \times 1}}$ are
\begin{equation}\label{eq20}
   {\mathbf{\gamma }}\left[ {nM + m} \right] = \left\{ {\begin{array}{*{20}{c}}
  {\sigma _d^2}&{{\text{others}}} \\ 
  0&\begin{gathered}
  \frac{N}{2} - {Q_L} \leqslant n \leqslant \frac{N}{2} + {Q_L}, \hfill \\
  \frac{M}{2} - L \leqslant m \leqslant \frac{M}{2} + L \hfill \\ 
\end{gathered}  
\end{array}} \right.,
\end{equation}
where $\sigma _d^2$ represents data symbol power. Lastly, due to ${{\mathbf{z}}_{\text{mod} }} = {\mathbf{F}}{{\mathbf{E}}_{{\text{mod}}}}{\mathbf{s}} $ can be rewritten as
\begin{equation}\label{eq21}
    {{\mathbf{z}}_{\text{mod} }} = \sum\limits_{l' = 0}^L {{\mathbf{F}}{\text{diag}}\left\{ {{{\mathbf{\Pi }}^{l'}}{\mathbf{s}}} \right\}{{\mathbf{e}}_{{\text{mod}},l'}}},
\end{equation}
where ${\mathbf{\Pi }}$ is the cyclic shift matrix, and ${{\mathbf{e}}_{{\text{mod}},l'}} = {\mathbf{G}}{{\mathbf{h}}_{l'}}$ denotes the modeling error corresponding to the $l'$th path. As ${\mathbf{s}}$ and ${{\mathbf{e}}_{{\text{mod}},l'}}$ are independent, the covariance matrix of ${\mathbf{E}}{{\mathbf{z}}_{\rm{mod}}}$ is given by
\begin{equation}\label{eq22}
    \begin{aligned}
   &\mathbb{E}\left\{ {\sum\limits_{l' = 0}^L {{\mathbf{EF}}{\text{diag}}\left\{ {{{\mathbf{\Pi }}^{l'}}{\mathbf{s}}} \right\}{{\mathbf{e}}_{\rm{mod ,}l'}}{\mathbf{e}}_{\rm{mod ,}l'}^H{\text{diag}}{{\left\{ {{{\mathbf{\Pi }}^{l'}}{\mathbf{s}}} \right\}}^H}{{\mathbf{F}}^H}{{\mathbf{E}}^H}} } \right\} \\ 
   &= \sum\limits_{l' = 0}^L {{\mathbf{EF}}\mathbb{E}\left\{ {{\text{diag}}\left\{ {{{\mathbf{\Pi }}^{l'}}{\mathbf{s}}} \right\}{{\mathbf{R}}_{{{\mathbf{e}}_{\rm{mod,}l'}}}}{\text{diag}}{{\left\{ {{{\mathbf{\Pi }}^{l'}}{\mathbf{s}}} \right\}}^H}} \right\}{{\mathbf{F}}^H}{{\mathbf{E}}^H}}  \\ 
   &= \sum\limits_{l' = 0}^L {{\mathbf{EF}}\left( {\left( {{{\mathbf{\Pi }}^{l'}}{\mathbf{F\Phi }}{{\left( {{{\mathbf{\Pi }}^{l'}}{\mathbf{F}}} \right)}^H}} \right) \odot {{\mathbf{R}}_{{e_{\rm{mod,}l'}}}}} \right){{\mathbf{F}}^H}{{\mathbf{E}}^H}},\\ 
    \end{aligned}
\end{equation}
where ${{\mathbf{R}}_{{{\mathbf{e}}_{\rm{mod,}l'}}}} = \mathbb{E}\left\{ {{{\mathbf{e}}_{\rm{mod,}l'}}{\mathbf{e}}_{\rm{mod,}l'}^H} \right\} = {\mathbf{G}}{{\mathbf{R}}_{{\mathbf{h}},l'}}{\mathbf{G}}$ is the variance of \({{\mathbf{e}}_{\rm{mod,}l'}}\). ${\mathbf{\Phi }} = {{\mathbf{\Phi }}_d} + {{\mathbf{\Phi }}_p}$ is a diagonal matrix consisting of data and pilot power, where ${{\mathbf{\Phi }}_p}$ is a diagonal matrix with a value of $\sigma _p^2$ only at the location of the pilot. $\sigma _p^2$ is pilot symbol power.

Combining (\ref{eq17}) with (\ref{eq22}), the proof of (\ref{eq13}) is completed.

 Then, by substituting (\ref{eq9}) into formula ${\bf{\tilde c}} = {\bf{c}} - {\bf{\hat c}}$, we have
 \begin{equation}\label{eq23}
 \begin{aligned}
  {\bf{c}} - {\bf{\hat c}} &= {\bf{c}} - \sqrt {{\xi _o}{\xi _i}} {\bf{D}}\left( {\sqrt {{\xi _o}{\xi _i}} {\bf{E}}{{\bf{A}}_p}{\bf{c}} + {\bf{Ez}}} \right) \cr 
   &= {\bf{Kc}} + \sqrt {{\xi _o}{\xi _i}} {\bf{DEz}}, \cr
 \end{aligned}
 \end{equation}
 where $\bf{D}$ and $\bf{K}$ are given in (\ref{eq_12}). Then, we obtain
 \begin{equation}\label{eq24}
 \begin{aligned}
  {{\bf{R}}_{{\bf{\tilde c}}}} &= {\mathbb E}\left\{ {{\bf{Kc}}{{\bf{c}}^H}{{\bf{K}}^H}} \right\} + {\mathbb E}\left\{ {{\xi _o}{\xi _i}{\bf{DEz}}{{\bf{z}}^H}{{\bf{E}}^H}{{\bf{D}}^H}} \right\} \cr 
   &= {\bf{K}}{{\bf{R}}_c}{{\bf{K}}^H} + {\xi _o}{\xi _i}{\bf{DE}}{{\bf{R}}_z}{{\bf{E}}^H}{{\bf{D}}^H}.
\end{aligned}
 \end{equation}
After that, the MSE for channel estimation is expressed as
\begin{equation} \begin{aligned}\label{eq25}
{\text{MS}}{{\text{E}}_{{\text{CE}}}} &= \frac{{\mathbb{E}\left\{ {{{\left( {{\mathbf{h}} - {\mathbf{\hat h}}} \right)}^H}\left( {{\mathbf{h}} - {\mathbf{\hat h}}} \right)} \right\}}}{{MN\left( {L + 1} \right)}}\\
 &= \frac{{\mathbb{E}\left\{ {{\bf{e}}_{\bmod }^H{{\bf{e}}_{\bmod }} + {{\left( {{\bf{\bar h}} - {\bf{\hat h}}} \right)}^H}\left( {{\bf{\bar h}} - {\bf{\hat h}}} \right)} \right\}}}{{MN\left( {L + 1} \right)}},
\end{aligned}\end{equation}
where ${{\bf{\bar h}}}$, ${{\bf{\hat h}}}$ denotes the channel vector modeled by GCE-BEM and the estimated channel vector, respectively. ${{\mathbf{e}}_{\text{mod}}} = {\mathbf{h}} - {\mathbf{\bar h}}$ is the modeling error vector. 

According to \cite{ref5}, the MSE of the BEM modeling error $\mathbb{E}\left\{ {{\mathbf{e}}_{\text{mod} }^H{{\mathbf{e}}_{\text{mod} }}} \right\} = \sum\limits_{l' = 0}^L {{\text{Trace}}\left\{ {{\mathbf{G}}{{\mathbf{R}}_{{\mathbf{h}},l'}}} \right\}}  $.
Subsequently, substituting  ${\mathbf{h}} = \left( {{{\mathbf{I}}_{L + 1}} \otimes {\mathbf{B}}} \right){\mathbf{Pc}}$ into (26), the MSE of the estimation error is given by
\begin{equation}\label{eq27}
   {\text{MS}}{{\text{E}}_{{\text{CE}}}} = \frac{{{\text{Trace}}\left\{ {{\mathbf{\Omega }}{{\mathbf{R}}_{{\mathbf{\tilde c}}}}{{\mathbf{\Omega }}^H}} \right\} + \sum\limits_{l' = 0}^L {{\text{Trace}}\left\{ {{\mathbf{G}}{{\mathbf{R}}_{{\mathbf{h}},l'}}} \right\}} }}{{MN(L + 1)}}.
\end{equation}
 \subsection{Proof of Theroem \ref{theorem2}}\label{appendix_2}
According to (\ref{eq14}), the $i$th symbol in the DD domain is defined as
\begin{equation}\label{eq28}
\begin{aligned}
{{{\mathbf{\hat x}}}_{d}}\left[ i \right] &= \sqrt {{\xi _o}{\xi _i}} {{\mathbf{f}}_i}{{\mathbf{G}}_t}{{\mathbf{H}}_t}{{\mathbf{F}}^H}{{\mathbf{x}}_{d}} + {{\mathbf{f}}_i}{{\mathbf{G}}_t}{\mathbf{n}}\\
&= \underbrace {{\mathbf{T}}\left[ {i,i} \right]{{\mathbf{x}}_{d}}\left[ i \right]}_{{\text{Desired signal}}} + \underbrace {\sum\limits_{j = 0,j \ne i}^{NM} {{\mathbf{T}}\left[ {i,j} \right]} {{\mathbf{x}}_{d}}\left[ j \right]}_{{\text{Intercarrier and InterDoppler interference}}} + \underbrace {{\mathbf{n'}}\left[ i \right]}_{{\text{Noise}}},
\end{aligned}
\end{equation}
where ${\mathbf{n'}} = {\mathbf{F}}{{\mathbf{G}}_t}{\mathbf{n}} \in {\mathbb{C}^{NM \times 1}}$ is the noise at the receiver, ${{\mathbf{f}}_i}$ is the $i$th row of the ${\mathbf{F}}$ matrix.
Then, the average power of ${{{\mathbf{\hat x}}}_{d}}\left[ i \right]$ is given by
\begin{equation}\label{eq29}
    \begin{aligned}
  \mathbb{E}\left\{ {{{\left| {{{{\mathbf{\hat x}}}_{d}}\left[ i \right]} \right|}^2}} \right\} &= \mathbb{E}\left\{ {\left( {{{{\mathbf{\hat x}}}_{d}}\left[ i \right]} \right){{\left( {{{{\mathbf{\hat x}}}_{d}}\left[ i \right]} \right)}^H}} \right\} \hfill \\
   &= {\xi _o}{\xi _i}\mathbb{E}\left\{ {{{\mathbf{f}}_i}{{\mathbf{G}}_t}{{\mathbf{H}}_t}{{\mathbf{F}}^H}{{\mathbf{x}}_d}{\mathbf{x}}_d^H{\mathbf{FH}}_t^H{\mathbf{G}}_t^H{\mathbf{f}}_i^H} \right\} \\
   &+ \mathbb{E}\left\{ {{{\mathbf{f}}_i}{{\mathbf{G}}_t}{\mathbf{n}}{{\mathbf{n}}^H}{\mathbf{G}}_t^H{\mathbf{f}}_i^H} \right\} \hfill \\
   &= {\xi _o}{\xi _i}{{\mathbf{f}}_i}{{\mathbf{G}}_t}\left( {{{\mathbf{H}}_t}{{\mathbf{R}}_{{{\mathbf{s}}_d}}}{\mathbf{H}}_t^H + {{\mathbf{R}}_{\mathbf{n}}}} \right){\mathbf{G}}_t^H{\mathbf{f}}_i^H \hfill \\  
   &= {\xi _o}{\xi _i}\sigma _d^2{\mathbf{T}}\left[ {i,i} \right] \hfill ,
    \end{aligned}
\end{equation}
Consequently, the SINR of the $i$th symbol can be calculated as
\begin{equation}\label{eq30}
    {\text{SINR}}\left[ i \right]{\text{ = }}\mathbb{E}\left\{ {\frac{{{{\left| {{\mathbf{T}}\left[ {i,i} \right]{{\mathbf{x}}_{d}}\left[ i \right]} \right|}^2}}}{{{{\left| {\sum\limits_{j = 0,j \ne i}^{NM} {{\mathbf{T}}\left[ {i,j} \right]} {{\mathbf{x}}_{d}}\left[ j \right] + {\mathbf{n'}}\left[ i \right]} \right|}^2}}}} \right\},
\end{equation}
We further derive the SINR using a similar approach as in \cite{ref12}. The numerator component of equation (\ref{eq30}) can be straightforwardly deduced as $\mathbb{E}\left\{ {{{\left| {{\mathbf{T}}\left[ {i,i} \right]{{\mathbf{x}}_{d}}\left[ i \right]} \right|}^2}} \right\} = {\mathbf{T}}{\left[ {i,i} \right]^2}\sigma _d^2$, and the denominator term designated as $\mathbb{E}\left\{ {{{\left| {\sum\limits_{j = 0,j \ne i}^{NM} {{\mathbf{T}}\left[ {i,j} \right]{{\mathbf{x}}_{d}}\left[ j \right]}  + {\mathbf{n'}}\left[ i \right]} \right|}^2}} \right\} = \mathbb{E}\left\{ {{{\left| {{{{\mathbf{\hat x}}}_{d}}\left[ i \right]} \right|}^2}} \right\} - \mathbb{E}\left\{ {{{\left| {{\mathbf{T}}\left[ {i,i} \right]{{\mathbf{x}}_{d}}\left[ i \right]} \right|}^2}} \right\}$. According to (\ref{eq29}), (\ref{eq30}) can be simplified as
\begin{equation}\label{eq31}
    {\text{SINR}}\left[ i \right]{\text{ = }}\frac{{{\mathbf{T}}\left[ {i,i} \right]}}{{1 - {\mathbf{T}}\left[ {i,i} \right]}}.
\end{equation}    
The average BER of an OTFS system considering the effect of hardware impairments can be expressed as
\begin{equation}\label{eq32}
    {P_{OTFS}} = \frac{1}{{{N_{{\text{num}}}}}}\sum\limits_{i = 0}^{{N_{{\text{num}}}}} {{a_M}{\text{erfc}}\left( {\sqrt {{b_M}{\text{SINR}}\left[ i \right]} } \right)},
\end{equation}
To further simplify the summation term, define the function $\eta \left( x \right) = erfc\left( {\sqrt {{b_M}\frac{x}{{1 - x}}} } \right)$ with $x \in \left( {0,1} \right)$.

Therefore, leveraging the convexity of $\eta \left( x \right)$\cite{ref14} and Jensen's inequality, the lower bound of (\ref{eq32}) can be derived as
\begin{equation}\label{eq33}
   {P_{OTFS}} \geqslant {a_M}{\text{erfc}}\left( {\sqrt {\frac{{{b_M}\sum\limits_{i = 1}^{{N_{{\text{num}}}}} {\frac{{{\mathbf{T}}\left[ {i,i} \right]}}{N}} }}{{1 - \sum\limits_{i = 1}^{{N_{{\text{num}}}}} {\frac{{{\mathbf{T}}\left[ {i,i} \right]}}{N}} }}} } \right).
\end{equation}
The inequality holds equality only when the diagonal elements of the ${\mathbf{T}}$ matrix are all equal. The proof has been completed.
}


\begin{thebibliography}{1}
\bibliographystyle{IEEEtran}

\bibitem{ref1}
X. He, W. Yuan and P. Fan, "On the pilot-aided channel estimation for windowed OTFS with data interference in rapidly time-varying channels," {\it{IEEE Trans. Wireless Commun.}}, vol. 23, no. 11, pp. 16359-16374, Nov. 2024. 

\bibitem{ref2}
J. Wu, W. Yuan, Z. Wei, K. Zhang, F. Liu and D. W. K. Ng, "Low-complexity minimum BER precoder design for ISAC systems: A delay-Doppler perspective," {\it{IEEE Trans. Wireless Commun.}}. 


\bibitem{ref3}
R. Hadani et al., "Orthogonal time frequency space modulation," in {\it{Proc. IEEE Wireless Commun. Netw. Conf. (WCNC)}}, Mar. 2017, pp. 1-6. 

\bibitem{ref4}
P. Raviteja, K. T. Phan and Y. Hong, "Embedded pilot-aided channel estimation for OTFS in delay–Doppler channels," {\it{IEEE Trans. Veh. Technol.}}, vol. 68, no. 5, pp. 4906-4917, May 2019. 

\bibitem{ref5}
Y. Liu, Y. L. Guan and D. G. G., "Near-optimal BEM OTFS receiver with low pilot overhead for high-mobility communications," {\it{IEEE Trans. Commun.}}, vol. 70, no. 5, pp. 3392-3406, May 2022. 


\bibitem{ref6}
Y. Liu, Y. L. Guan and D. G. G., "Turbo BEM OTFS receiver with optimized superimposed pilot power," {\it{IEEE Trans. Commun.}}, vol. 72, no. 1, pp. 601-617, Jan. 2024. 

\bibitem{ref7}
A. Tusha, S. Doğan-Tusha, F. Yilmaz, S. Althunibat, K. Qaraqe and H. Arslan, "Performance analysis of OTFS under in-phase and quadrature imbalance at transmitter," {\it{IEEE Trans. Veh. Technol.}}, vol. 70, no. 11, pp. 11761-11771, Nov. 2021. 

\bibitem{ref8}
S. G. Neelam and P. R. Sahu, "Analysis, estimation and compensation of hardware impairments for CP-OTFS systems," {\it{IEEE Wireless Commun. Letters}}, vol. 11, no. 5, pp. 952-956, May 2022. 

\bibitem{ref9}
A. Singh, S. Sharma, M. Sharma, K. Deka and D. B. da Costa, "Autoencoder-based end-to-end OTFS system design with hardware impairments," {\it{IEEE Wireless Commun. Letters}}, vol. 13, no. 8, pp. 2285-2289, Aug. 2024.

\bibitem{ref10}
X. Li, J. Li, Y. Liu, Z. Ding and A. Nallanathan, "Residual transceiver hardware impairments on cooperative NOMA networks," {\it{IEEE Trans. Wireless Commun.}}, vol. 19, no. 1, pp. 680-695, Jan. 2020. 

\bibitem{ref11}
D. Tubail, B. Ceniklioglu, A. E. Canbilen, I. Develi and S. S. Ikki, "The effect of hardware impairments on the error bounds of Localization and maximum likelihood estimation of mm-Wave MISO-OFDM systems," {\it{IEEE Trans. Veh. Technol.}}, vol. 72, no. 3, pp. 4063-4067, March 2023. 

\bibitem{ref12}
Yuan-Pei Lin and See-May Phoong, "BER minimized OFDM systems with channel independent precoders," {\it{IEEE Trans. Signal Processing}}, vol. 51, no. 9, pp. 2369-2380, Sept. 2003. 

\bibitem{ref13}
A. Goldsmith, {\it{Wireless communications}}. Cambridge university press, 2005.

\bibitem{ref14}
Z. Li, C. Zhang, G. Song, X. Fang, X. Sha and D. Slock, "Chirp Parameter Selection for Affine Frequency Division Multiplexing with MMSE Equalization," \it{IEEE Trans. Commun.}.
\end{thebibliography}
\end{document}